\documentclass[12pt]{article}

\title{The quark-gluon Mixed Condensate calculated via Field Correlators}

\author{ A.Di Giacomo$^{(a)}$, Yu.A.Simonov$^{(b)}$\\
$^{(a)}$Dipartimento di Fisica ``Fermi" and INFN-Sezione di
Pisa,\\ Via Buonarroti, 2 -- Ed. C -- I-56127 Pisa, Italy\\
$^{(b)}$ State Research Center\\Institute of Theoretical and
Experimental Physics, \\ Moscow, 117218 Russia}

 \date{}

\newcommand{\beq}{\begin{eqnarray}}

 \newcommand{\eeq}{\end{eqnarray}}

\newcommand{\be}{\begin{equation}}

 \newcommand{\ee}{\end{equation}}

 \def\la{\mathrel{\mathpalette\fun <}}

\def\fun#1#2{\lower3.6pt\vbox{\baselineskip0pt\lineskip.9pt

\ialign{$\mathsurround=0pt#1\hfil ##\hfil$\crcr#2\crcr\sim\crcr}}}

\newcommand{{\SD}}{\rm SD}

\newcommand{\lan}{\langle}

\newcommand{\ran}{\rangle}

\begin{document}

\maketitle

\begin{abstract}

The quark-gluon mixed condensate $g\lan \bar q \sigma_{\mu\nu}
F_{\mu\nu} q\ran$ is calculated in the Gaussian approximation of
the Field Correlator Method.  In the large $N_c$ limit and for
zero mass quarks one obtains a simple result, $m^2_0 \equiv
\frac{g\lan \bar q \sigma_{\mu\nu} F_{\mu\nu} q\ran}{\lan \bar q
q\ran} = \frac{16\sigma}{\pi} $, where $\sigma$ is the string
tension. For a standard value $\sigma =0.18$ GeV$^2$ one obtains
$m^2_0= 1$ GeV$^2$ in   good agreement with the QCD sum rules
estimate $m^2_0 = (0.8\pm 0.2) $ GeV$^2$ and  the latest lattice
result $m^2_0\cong 1$ GeV$^2$.

\end{abstract}

\section{ Introduction}

The mixed quark-gluon condensate (QGC) is an important
characteristics of the nonperturbative QCD vacuum, which together
with the quark condensate $\lan \bar q q\ran$ signals the chiral
symmetry breaking. Moreover,  the QGC measures the average
interaction of the quark color-magnetic moment with the vacuum
fields, which is an important ingredient of the quark dynamics in
the vacuum (e.g. it is  this term which gives attraction of in
quark zero modes).

In the QCD sum rules  the QGC plays an important role \cite{1} and
the phenomenological analysis suggests the value of $m^2_0$ in the
range $m^2_0= (0.8\pm 0.2)$ GeV$^2$
 \cite{1}, see  \cite{2} for a
review. One should stress at this point that for  a nonzero quark
mass $m$
 the (diverging) perturbative part  should  be subtracted.

As will be seen below the resulting  nonperturbative dependence of
$m^2_0$ on $m$ is very weak in agreement with lattice data.
Lattice studies of QGC \cite{3},\cite{4},\cite{5} have not yet
converged to  a definite prediction. A problem there is the
extrapolation to zero quark mass and the quenched approximation.
In ref. \cite{4} the simulations are done in the quenched
approximation, the condensate is measured by use of staggered
quarks, and the result for $m^2_0$ is definitely larger than the
sum-rules value. Ref. \cite{5} uses an optimized version \cite{55}
of  domain wall fermions, which are better in principle for the
chiral limit, again in the quenched approximation. Their result is
$m^2_0= 1$GeV$^2$, which agrees with QCD sum rules. It is
therefore worthwhile to calculate QGC by a different
nonperturbative method.

In the framework of the Field Correlator Method (FCM) \cite{6} the
color-magnetic quark gluon interaction term $g\sigma_{\mu\nu}
F_{\mu\nu} $ enters essentially in the Fock-Feynman-Schwinger
Representation  (FFSR) of the quark propagator in the vacuum
background field \cite{7}. In particular the quadratic average of
this term defines the hyperfine $q\bar q$ interaction where   the
nonperturbative part is proportional to the field correlator $\lan
F_{\mu\nu}(x) F_{\rho\sigma}(0)\ran$ measured on the lattice
\cite{8}. Even more important the term $g\sigma_{\mu\nu}
F_{\mu\nu}$ is in the contribution to the  bound quark self-energy
\cite{9}, where it is of paramagnetic character, i.e. negative and
strongly decreases the masses of hadrons, putting them in
accordance with experimental data \cite{10}. Explicit correction
to the bound quark mass squared is \cite{9}
\be
\Delta m^2_q= -\frac{4\sigma}{\pi} \eta \label{1}\ee where
$\eta=\eta(mT_g)$ is a calculable function of the quark current
mass $m$, renormalized at the scale of 1 GeV. The function $\eta$
is given in \cite{9} and  in the Appendix below and for zero quark
mass is normalized to one: $\eta(0)=1$. We calculate  in the next
section the QGC, or rather the parameter $m^2_0$ in the same way,
as it was done in \cite{9} for $\Delta m^2_q$,  with the result
\be
m^2_0 =- 4 \Delta m^2_q = \frac{16\sigma}{\pi} \eta.
 \label{2}\ee
For $\sigma = 0.18 $ GeV$^2$ one obtains $m^2_0= 1 $ GeV$^2$ which
is  in  agreement with the lattice data \cite{5}, and with the QCD
sum rules estimate quoted above.

\section{Calculation of $m^2_0$}

We proceed in the Euclidean space-time and write
\be
\lan \bar q g \sigma_{\mu\nu} F_{\mu\nu} q\ran_{q,A}= tr \lan
g\sigma_{\mu\nu} F_{\mu\nu} (x) S_q (x,x) \ran_A= tr \lan S_q(x,x)
g\sigma_{\mu\nu}F_{\mu\nu}(x)\ran\label{3}\ee where $S_q(x,y)$ is
the Euclidean quark propagator, for which one can write using the
FFSR $$ S_q (x,y) =(m+\hat D)^{-1}_{x,y} = (m-\hat D)_x (m^2-\hat
D^2)^{-1}_{x,y}= $$
\be
(m-\hat D)\int^\infty_0 ds (Dz_{x,y} e^{-K} \Phi_z(x,y) P_F\exp
\int^s_0 \lambda(z(\tau)) d\tau .\label{4}\ee In (\ref{4}) the
following notations are used: $K=m^2s+\frac24
 \int^s_0 \dot z_\mu^2 d\tau$, $ D_\mu\equiv \partial_\mu -ig
 A_\mu, (Dz)_{x,y}$ is the path-integral measure for paths
 starting at $y$ and ending at the point $x$, $(Dz)_{x,y}
 =\lim_{N\to\infty} \prod^N_{n=1} (\frac{d^4\Delta
 z(n)}{(4\pi\varepsilon)^2}) \frac{d^4k}{(2\pi)^4}
 e^{ik(\sum^n_{n=1} \Delta z(n)-(x-y))}$, while $\Phi_z(x,y)$ is
 the phase factor (parallel transporter) along the path
 $z_\mu(\tau)$ $\Phi_z(x,y) = P_A\exp ig \int^x_y A_\mu dz_\mu$,
 with $P_A, P_F$ - the ordering operators of the matrices
 $A_\mu(z)$ and $\lambda(z)$, where $\lambda(z)$ is defined to
 be\footnote{the definition of  $\sigma_{\mu\nu}$ in (5) (as well
 as  in \cite{6},\cite{7}) differs from the standard definition in
 QCD sum rules, where enters $\frac12$ instead  of $\frac14$  in (5). Therefore one obtains additional factor 2
 in the definition
 of $m^2_0$ in (17).}
 \be
 \lambda(z) \equiv  g\sigma_{\mu\nu} F_{\mu\nu}(z) ; ~~
 \sigma_{\mu\nu} =\frac{1}{4i} (\gamma_\mu
 \gamma_\nu-\gamma_\nu\gamma_\mu).\label{5}\ee
 For what follows it will be advantageous to take in (\ref{5})
 $\lambda(z(\tau))= g(\tau) \sigma_{\mu\nu} F_{\mu\nu} (z(\tau))$,
 since the functional derivative $\frac{\delta}{\delta g(\tau)} $
 at $\tau\to 0$  or $\tau\to s$ inside the FFSR (\ref{4}) brings
 down
 additional factor $\lambda(y)$ or $\lambda(x)$. When one has
 $y=x$, as in (\ref{3}), then both contributions add, which
 formally is obtained by putting $g(0)=g(s)$. In this way one can
 rewrite (\ref{3}) as follows
\be
\lan \bar q(x) \lambda (x) q (x) \ran = tr \lan \lambda (x) S_q
(x,x) \ran = 2 tr \frac{\delta}{\delta g(0)} \lan S_q(x,x)\ran .
\label{6}\ee As the next step one can write the average $\lan
S_q(x,x)\ran$ in the form of cluster expansion \cite{6}
\be
\lan S_q(x,x)\ran = \lan (m-i\hat p) \int^\infty_0 ds e^{-K}
(Dz)_{xx} \exp \{ -\frac12 \int dv_{\lambda\rho} \int
dv_{\sigma\nu} \lan g F_{\lambda\rho} gF_{\sigma \nu}\ran
\}\label{7}\ee where only the contribution of the lowest cumulant
$\lan FF\ran$ is retained in accordance with estimates \cite{11},
and the nonabelian Stokes theorem is used to express $A_\mu$
through $F_{\mu\nu},$ with the notation $$ dv_{\lambda \rho} =
ds_{\lambda\rho}- i\sigma_{\lambda\rho} d\tau,~~ gF_{\lambda \rho}
dv_{\lambda\rho} =$$ \be= gF_{\lambda\rho} (u) ds_{\lambda\rho}
(u) -ig (\tau) \sigma_{\lambda\rho}
F_{\lambda\rho}(z(\tau))\label{8}\ee and $ds_{\lambda\rho} $ is
the element of the area of the surface enclosed by the contour
$z_\mu(\tau), z_\mu(0)= z_\mu(s) =x_\mu$. Performing
differentiation in (\ref{6}) one gets $$ \lan \bar q \lambda
q\ran= 2g^2 \sigma_{\mu\nu} \sigma_{\lambda\rho} \int^\infty_0 ds
(Dz)_{xx} e^{-K} (m-i\hat p) \int^s_0 d\tau \lan F_{\lambda\rho}
(u(\tau)) F_{\mu\nu} (x)\ran\times $$\be \times \exp \{
-\frac{g^2}{2} \int dv_{\lambda\rho} \int dv_{\sigma\nu} \lan
F_{\lambda \rho} F_{\sigma \nu}\ran\}.\label{9}\ee

Using the identities \cite{7}
\be
(Dz)_{xx} = (Dz)_{xu} d^4u (Dz)_{ux}, \int^\infty_0 ds \int^s_0
d\tau f(s, \tau)= \int^\infty_0 ds \int^\infty_0 d\tau f(s+\tau,
\tau)\label{10}\ee where $f(s,\tau)$ is an arbitrary function, one
has
\be
\lan \bar q \lambda q \ran= 2 \sigma_{\mu\nu} \sigma_{\lambda\rho}
\int \lan G(x,u) S_q(u,x) \ran D^{(2)}_{\lambda\rho,\mu\nu} (u-x)
d^4(u-x).\label{11}\ee Here we have defined as in \cite{6}
\be
D^{(2)}_{\lambda\rho, \mu\nu} (z) \equiv (\delta_{\lambda\mu}
\delta_{\rho\nu}- \delta_{\lambda\nu} \delta_{\rho\mu} ) D(z)
+\frac12 (\partial_\lambda z_\mu \delta_{\rho\nu} +\partial_\rho
z_\nu \delta_{\lambda\mu} -\partial_\lambda z_\nu
\delta_{\rho\mu}-\partial_\rho z_\mu \delta_{\lambda\nu} )D_{1}
(z)\label{12}\ee and
\be
G(x,u)=\int^\infty_0 d\tau e^{-K} (Dz)_{xu}\exp \{ -\frac12 \int
dv_{\lambda\rho} \int dv_{\sigma\nu} \lan g F_{\lambda\rho}
gF_{\sigma \nu}\ran \}.\label{13} \ee Note that $G_0(x,u)$ and
$S_q(u,x)$ share common factors depending on a piece of common $z$
between $u_\mu$ and $x_\mu$ and in general do not factorize.

At this point we shall use the properties of the correlators
$D(z), D_1(z)$ found on the lattice \cite{8},  in the quenched
case one has
\be
D(z) \cong 3 D_1(z) = D(0) \exp (-|z|\delta),~~ \delta \equiv
1/T_g \approx 1~{\rm GeV}.\label{14}\ee

Analytic calculations based on the gluelump spectrum [12] suggest
even larger value, $\delta\approx 1.4\div 1.5$ GeV. The string
tension $\sigma$ can be expressed through $D(z)$ (the correction
due to higher correlators is limited by the Casimir scaling
arguments to a few percent \cite{11})
\be
\sigma=\frac12 \int D(z) d^2z.\label{15}\ee Since the distance
$|u-x|$ is of the order of $T_g$, we can now use the argument of
the small $T_g$ limit (large $\delta)$ for  the constant $\sigma$
to factorize the product $\lan G(x,u)S_q(u,x)\ran$ as follows
\be
\lim_{T_g\to 0} \lan G(x,u) S_q(u,x)\ran \cong G_0 (x-u) \lan
S_q(x,x)\ran.\label{16}\ee

This approximation is equivalent to the expansion in the parameter
$\xi \equiv \sigma T^2_g\ll 1$. As the result one obtains the
following representation for the ratio
\be
m^2_0\equiv  2\frac{\lan \bar q\lambda q\ran}{\lan \bar q q\ran} =
4 \sigma_{\mu\nu} \sigma_{\lambda\rho} \int G_0 (z)
D^{(2)}_{\lambda\rho, \mu\nu} (z) d^4 z\label{17}\ee $G_0(z)$ is
easily calculated using (\ref{13}) to be the free propagator of
the scalar quark with mass $m$,
\be
G_0(z) = \frac{mK_1(m|z|)}{4\pi^2|z|}\label{18}\ee where $K_1$ is
the McDonald function, and $m$ is the current (pole) quark mass
normalized at 1 GeV.

Taking into account that\footnote{note the misprint in Eq.(15) of
[9], where coefficients of $D, D_1$ differ from those in
(\ref{19}). Nevertheless the final result in Eq.({29}) of \cite{9}
is the same as in our Eq. (\ref{1}) due to the relation
$D_1\approx \frac13 D$ \cite{8} valid for the quenched case,
considered here, whereas in the unquenched  case one obtains
instead of (1): $\Delta m^2_q{(m\to 0)}=- 3\int^\infty_0 zdz
(D+D_1) \cong -\frac{3}{\pi}\sigma$.}

 \be\sigma_{\mu\nu}
\sigma_{\lambda\rho} D^{(2)}_{\lambda\rho,\mu\nu} (z) = 6 (D(z) +
D_1(z))\label{19} \ee one obtains for $m^2_0$
\be
m^2_0= 12m \int^\infty_0 z^2 dz K_1(mz) (D(z)+ D_1(z))\label{20}
\ee or, with the help of (\ref{14}),
\be
m_0^2\cong 16 m \int^\infty_0 z^2dz K_1(mz) D(z) =
\frac{16\sigma}{\pi} \varphi(m/\delta)\label{21}\ee where we have
defined
\be
\varphi(m/\delta) \equiv m \delta^2 \int^\infty_0 z^2dz K_1 (mz)
\exp(-\delta z);  \varphi(0)=1.\label{22}\ee

It is easy to see with the help of (\ref{15}) that in the limit of
small quark mass, $m\to 0$, one obtains for $\sigma=0.18$
GeV$^2$(in the quenched case)
\be
m^2_0 ( m\to 0) =\frac{16}{\pi} \sigma= 0.92 {\rm~
GeV}^2.\label{23} \ee

It is appropriate at this point to discuss the accuracy of our
result (\ref{23}). The main uncertainty  appears in expressions
(\ref{14}), (\ref{15}) and (\ref{16})  and we consider  the
accuracy of the  corresponding approximations point by point.

The lattice calculations [9] of $D(z)$  and $D_1(z)$ define the
amplitudes $A, A_1$ and slopes $\delta, \delta_1$; the first ones
are  reabsorbed in the  value of $\sigma$, while the latter are
equal with accuracy of few (1-2) percent to the value given in
(\ref{14}). The approximation of (\ref{15}) reduces to the neglect
of higher correlators, contributing to the observed  string
tension $\sigma$. This accuracy was tested in [12] using the
Casimir scaling and is also of the order of few percent. The
largest possible error may come from the replacement (\ref{16}),
where one can use the fact that the integral over  $d^4(u-x)$ in
(\ref{11}) is taken with the weight $D^{(2)}(u-x)$. The latter is
exponentially decreasing at the distance $1/\delta$, while the
range of $G(x,u)$ is defined by the confining exponent in
(\ref{13}), which produces the effective quark mass, computed
through $\sigma$ and equal to 0.35 GeV for the lowest state (see
[11] for references and explicit calculations). Introducing this
mass instead of $m$ in (\ref{18}), (\ref{20}), (\ref{21}) one
obtains $\varphi\approx 0.75\div 0.8$, and using (\ref{21}) one
comes to the conclusion that  $m^2_0$ is in the range $0.7$
GeV$^2\la m_0^2\la 1$ GeV$^2$. This range lies very close to the
limits predicted in the QCD sum rules.

 The
explicit analytic form of $\varphi(x)$ was obtained in \cite{9}
and is given here in Appendix. For $\delta=1$ GeV, and $m=0.175$
GeV, 1.7 GeV and 5 GeV one obtains respectively $\varphi = 0.88,
0.234 $ and 0.052.

The resulting value of $m^2_0$ (\ref{23}) is in  agreement with
the QCD sum rule estimates \cite{2},  and with  the lattice
evaluation of $m^2_0$, namely  $m^2_0 \approx 1$ GeV$^2$ in
\cite{5}. One should note, that there is a large perturbative
contribution to $m^2_0$ for nonzero quark mass $m$ proportional to
$m\Lambda^2_{UV}\sim m/a^2$, which should be subtracted to get
agreement with purely nonperturbative result (\ref{23}).

On the other hand  the purely nonperturbative behaviour of $m^2_0$
as a function of the quark mass $m$, or rather the ratio
$t=m/\delta$ is given in appendix, Eq. (A.6), \be m^2_0 (t)
=\frac{16\sigma}{\pi} (1+ t^2 (4-3\ln \frac{2}{t}) +
O(t^4)).\label{24} \ee The values $m^2_0(t)$ obtained from
(\ref{24}) agree well with the lattice measured values in \cite{5}
for  $m a>0$. Indeed   for three values of $ma$, $ma=0.05; 0,1 $
and  0.15 one obtains from (\ref{24}) taking $\sigma =0.18$
GeV$^2$, and  $a^{-1}=1,979$ GeV \cite{5}, $ m^2_0 = 0.434; 0.393$
and  $0.342$ GeV$^2$ respectively. This should be compared with
the values $m^2_0(ma)$ measured in \cite{5} and equal to 0.371;
0.311 and 0.290 GeV$^2$. At the same time the limiting
extrapolated value $m^2_0 (ma=0) \approx 1$ GeV$^2$ obtained in
\cite{5},  agrees with the  theoretical one, given by Eq.
(\ref{24}), $m_0^2(ma=0,~theory)=1$ GeV$^2$. One should have in
mind, that chiral quark mass corrections present in both the quark
condensate and the QGC are cancelled in the ratio $m_0^2$ to the
leading order in $\sigma T^2_g$, so the remnant $ma$ dependence in
$m^2_0$ comes from quadratic terms in (\ref{24}) and linear
perturbative terms  mentioned above.

Recently a study of thermal dependence of $m^2_0(T)$ has been
reported in \cite{13}, where $m^2_0$ was found almost independent
of $T$ up to $T=T_c$. This is in general agreement  with our
expression (23), since $\sigma$ is roughly constant in that
region, but more detailed check of behaviour near $T_c$ is
desirable.

 Summarizing, we have obtained a simple nonperturbative
estimate for the ratio of condensates, which is in a reasonable
agreement with the QCD sum rule results, and lattice results in
\cite{5} for nonzero $ma$ and zero $ma$ limit.

The authors are grateful to D.V.Antonov for many useful
discussions and collaboration at the first stage of this work.

The partial support of the INTAS grants 00-110 and 00-366 is
gratefully acknowledged.

The work  of  Yu.S. was supported by the Federal Program of the
Russian Ministry of Industry, Science and Technology No
40.052.1.1.1112.

\newpage

 \setcounter{equation}{0}

\renewcommand{\theequation}{A.\arabic{equation}}

\begin{center}

{\bf Appendix }\\

 \vspace{0.5cm}

\end{center}

The function $\varphi(t), t\equiv m/\delta$, defined in Eq.(22)
can be written as (note the difference in definition here and in
\cite{9})
\be
\varphi(t)= t\int^\infty_0 z^2 dzK_1(tz) e^{-z}\label{a.1} \ee
where $K_1$ is the McDonald function, $K_1(x) (x\to 0) \approx
\frac{1}{x}$,  so that for $t=0$ one obtains
\be
\varphi(0)=1.\label{a.2}\ee For $t>0$ the integration in
(\ref{a.1}) yields two different forms; e.g. for $t<1$,
\be
\varphi(t) =- \frac{3t^2}{(1-t^2)^{5/2}} \ln
\frac{1+\sqrt{1-t^2}}{t}+ \frac{1+2t^2}{(1-t^2)^2}\label{a.3}\ee
while for $t>1 $ one has instead,
\be
\varphi(t) =- \frac{3t^2}{(t^2-1)^{5/2}}\arctan
(\sqrt{t^2-1})+\frac{1+2t^2}{(1-t^2)^2}.\label{a.4}\ee For large
$t$ one has the following limiting behaviour, \be \varphi(t)
=\frac{2}{t^2} - \frac{3\pi}{2 t^3} + O(\frac{1}{t^4}).\label{a.5}
\ee

For small $t$ one obtains expanding the r.h.s. of (\ref{a.3})
\be
\varphi(t) =1 +t^2 (4-3\ln \frac{2}{t}) + t^4 (\frac74 -
\frac{15}{2} \ln \frac{2}{t} )+ O(t^6). \label{a.6} \ee
 Some numerical
values are useful in applications. $$\varphi(0.175) \cong 0.88,
~~\varphi (1.7) \cong 0.234, \varphi (5) \cong 0.052$$

\end{document}